\documentclass[aps,eqsecnum,preprint,floats,nofootinbib,showpacs]{revtex4}
\usepackage{amsmath,amssymb,amsthm}
\usepackage{graphicx}
\usepackage{hyperref}
\def\be{\begin{eqnarray}}
\def\en{\end{eqnarray}}
\def\non{\nonumber}
\def\la{\langle}
\def\ra{\rangle}

\def\prl{{ Phys. Rev. Lett.}~}

\def\bi{\bibitem}

\def\ua{\uparrow}
\def\da{\downarrow}
\begin{document}

\title{\large \bf  A Difference Operator Approach to Quantum Random Walks: 
Parseval Identity, Krawtchouk Matrices, and Hermite Limits }

\author{ \bf  Chien-Wen Hwang\footnote{
t2732@mail.nknu.edu.tw}}

\affiliation{\centerline{Department of Physics, National Kaohsiung Normal University,} \\
\centerline{Kaohsiung, Taiwan 824, Republic of China}
 }


\begin{abstract}
We introduce a discrete difference operator $D_k$ to study the 
one-dimensional quantum random walk (QRW) with the Hadamard coin. 
Explicit combinatorial expressions are obtained for the probability 
amplitudes $a(n,k)$ and $b(n,k)$, which encode the final step 
direction and carry alternating signs that reflect the merging of 
leftward steps. Removing these signs and the coin-state distinction 
recovers the classical binomial distribution. The symmetric and 
antisymmetric combinations $a\pm b$ are shown to coincide with 
diagonal and sub-diagonal entries of the Krawtchouk matrix. Using 
cross identities among Krawtchouk matrix elements, we prove by 
induction that the amplitudes satisfy a Parseval identity 
$\sum (a^2+b^2)=2^n-1$, establishing probability conservation in 
the Krawtchouk formulation. The operator $D_k$ acts as a discrete 
Hermite polynomial generator: the ratios $h_m = \binom{n}{k}^{-1} 
D_k^m \binom{n}{k}$ admit explicit closed forms and converge to 
Hermite polynomials in the continuous limit. At the discrete level, 
$D_k$ connects successive Krawtchouk matrices and acts as a 
coherence generator and a raising operator. The $h_m$ quantify 
the position-dependent degree of quantum 
interference on both halves of the distribution, either 
individually (for $x \ge 0$) or through an inverse-Pascal 
combination of several $h_m$ (for $x < 0$), and the 
ballistic $O(n^2)$ scaling of the variance emerges from the 
superposition of all excited states. Numerical illustrations for 
$n=6$ and $n=10$ corroborate the analytical results.
\end{abstract}
\maketitle %

\section{Introduction}
Among the many quantum algorithms, quantum walk algorithms have 
attracted considerable attention due to their independence from 
specific quantum hardware implementations and their role as a 
universal computational primitive~\cite{childs2009, lovett2010}. The 
concept of the quantum random walk (QRW) was first introduced in 
1993~\cite{aharonov1993}. As an extension of the classical random 
walk (RW), it has since had a unique impact on quantum computation, 
quantum information, and spatial search problems, offering, in 
particular, exponential speedups over classical RW for certain 
algorithmic tasks \cite{ambainis2001, kempe2003}. 

Generally speaking, QRW models fall into two classes: discrete-time 
\cite{DTQW} and continuous-time \cite{CTQW}. In the former, the 
dynamics are driven by the repeated action of a shift operator, whereas 
in the latter they are governed by a time-independent Hamiltonian. At 
the heart of most discrete-time QRW models lies a distinctive feature: 
the walker possesses an internal coin state that records the direction of 
the last step. This quantum memory, absent in classical diffusive motion, 
is responsible for the ballistic spread of the probability distribution and 
for the rich interference patterns that characterize quantum walks. 
Different choices of the coin operator can be used to simulate 
different physical systems and can also enhance the speed of the 
walk~\cite{faster}. The most common choices and their properties are 
summarized below:
\begin{itemize}
\item[(i)] Hadamard coin, providing a prototypical walk that exhibits 
interference phenomena and ballistic spreading.
\item[(ii)] Grover coin, commonly used in quantum search algorithms;
it can locate targets more rapidly on graphs.
\item[(iii)] Parameterized SU(2) coin, whose parameters can be tuned to 
simulate physical phenomena such as those with topological 
characteristics.
\end{itemize}

There also exist several different approaches to analyzing the 
probability distribution of the QRW. Meyer~\cite{meyer1996} first 
built on the combinatorial and path-integral methods to express the QRW 
amplitudes. Ambainis et al.~\cite{ambainis2001} 
subsequently  expressed them in terms of Jacobi polynomials and used Fourier 
analysis to extract the long-time asymptotics of these amplitudes, which are 
described by Bessel functions. Grimmett et al.~\cite{grimmett2004} 
later developed a probabilistic approach based on weak convergence, 
which simplified and extended the one-dimensional analysis.

Despite extensive research, the combinatorial structure underlying the 
probability amplitudes of quantum walks has remained relatively 
unexplored. Most treatments start from a unitary coin operator (such as 
the Hadamard matrix) and compute amplitudes recursively, leaving the 
explicit form of the amplitudes as functions of the step number $n$ and 
the position $k$ somewhat opaque. Konno~\cite{konno2002} made an 
early step toward explicit combinatorial expressions using path counting 
and generating functions. A notable further advance is the work of 
Feinsilver and Kocik \cite{feinsilver2004, feinsilver2007}, who recognized 
that Krawtchouk polynomials appear naturally in the spectral analysis of 
the Hadamard walk. However, the direct construction of the walk's 
amplitudes from elementary combinatorial operations, and the 
identification of their algebraic structure with Krawtchouk matrix 
elements, has not been fully achieved.

In this paper, we start from a purely combinatorial operation: 
constructing the probability amplitudes $a(n,k)$ and $b(n,k)$ as a 
convolution of two binomial coefficients, where $n$ and $k$ denote 
the total number of steps and the number of leftward steps, 
respectively. This construction makes the underlying combinatorial 
structure of the QRW explicit; throughout the derivation, the 
classical binomial distribution served as a recurring benchmark, 
requiring that the quantum amplitudes reduce to the classical 
result upon removing interference. Building on these amplitudes, 
we define a combinatorial difference operator $D_k$ that identifies 
$a \pm b$ with diagonal and sub-diagonal entries of the Krawtchouk 
matrix and gives rise to the ratios $h_m = \binom{n}{k}^{-1} 
D_k^m \binom{n}{k}$, which serve as discrete analogues of the 
Hermite polynomials $H_m$ and form the basis of our analysis of the 
probability distribution, both in the right half ($x \ge 0$) and, via a 
difference structure, in the left half ($x \le 0$).

Although any complete basis can be used to expand the QRW 
probability distribution, the choice of Hermite functions is 
physically motivated: the QRW distribution is characterized by 
ballistic peaks at the extremities and a dip at the origin---a 
shape naturally captured by higher-order Hermite functions. 
Moreover, the generator of the Hermite polynomials is the 
continuous counterpart of $D_k$, making the Hermite expansion 
not merely a mathematical convenience, but a reflection of the 
underlying physics.

The remainder of this paper is organized as follows. In Section~II we review the 
necessary background on one-dimensional QRW, and drive the probability 
amplitude in datrils; the procedure by which these amplitudes reduce to the 
RW probabilities is also discussed. In Section~III we connect 
the combinatorial amplitudes $a(n,k)$ and $b(n,k)$ to the elements of Krawtchouk 
matrix and prove a Parseval idenity for the QRW by induction, establishing the 
conservation of probability. Section~IV introduces a modified difference operator 
$D_k$ and shows that it acts as a discrete analogue of the Hermite polynomial 
generator. Explicit expressions for the ratios $h_m(n,k)$ are derived, and their 
continuous limit is shown to recover the standard Hermite 
polynomials $H_m(\xi)$. In Section~V, we reveal that $D_k$ connects successive 
Krawtchouk matrices and acts as a coherence generator, converting classical 
binomial probabilities and lower-order quantum amplitudes alike 
into higher-order interfering amplitudes. The $h_m$ are shown to 
govern the contributions to the variance on both halves of the 
distribution; the $O(n^2)$ ballistic scaling emerges from the 
superposition of all excited states, which is the discrete 
counterpart of the Hermite-function oscillations that shape the 
QRW probability density in the continuous limit. Finally, a conclusion is given 
in Section~VI with open questions and directions for higher-dimensional 
generalizations.


\section{From QRW to RW}
Since quantum mechanics constitutes the underlying architecture of our 
physical world, subsuming classical behavior as a limiting case, we treat  
QRW as the fundamental starting point for the 
study of random walks. As mensioned in Ref.  \cite{kempe2003}, the so 
called discrete-time QRW in one dimension is defined on the Hilbert 
space $\mathcal{H} = \mathcal{H}_C \otimes \mathcal{H}_P$, where 
$\mathcal {H}_C$ and $\mathcal{H}_P$ spanned by the basis coin states 
$\{|R\ra, |L\ra\}$ and the basis position states $\{|i\ra : i \in {\bf Z}\}$, 
respectively.  We denote the coin states by $|R\ra$  and $|L\ra$ , 
corresponding respectively to the spin states $|\ua \ra$ and $|\da\ra$ 
in the standard quantum information notation.

The  $n$-step QRW is realized by the unitary transformation $U^n$ 
acting on the Hilbert space $\mathcal{H}$, where
\be
U=S\cdot (C\otimes I),
\en
is the unitary evolution operator for a single step.
Here, $S$ is the shift operator of position $|i\ra$ based the coin states 
$|L\ra=(0,1)^T$ and $|R\ra=(1,0)^T$:
\be
S=|L\ra \la  L | \otimes \sum_i |i-1\ra \la i|+|R\ra\la R|\otimes \sum_i |i+1\ra \la i|,
\en
$C$ is coin-flip operator to the left ($|L\ra$) or right ($|R\ra$) state , and 
$I$ is the identity operator of position space. The coin state determines 
the direction of each step and serves as an internal memory that records 
the direction of the immediately preceding step, 
a concept first introduced by Aharonov, Davidovich, and Zagury  
\cite{aharonov1993}. Consequently, after $U^n$ acts on the initial state, 
the probability amplitude of the walker at any position can be decomposed 
into two parts: one arising from paths ending with a leftward step, and the 
other from paths ending with a rightward step. In this work, we adopt the 
commonly used Hadamard coin:
\be
C\to H=\frac{1}{\sqrt{2}} \left( \begin{array}{cc} 1 & 1 \\ 1 & -1 \end{array} \right),
\en
and have the superposition states
 \be
 H|R\ra=\frac{1}{\sqrt{2}}(|R\ra+|L\ra), ~~~     H|L\ra=\frac{1}{\sqrt{2}}(|R\ra-|L\ra).
 \en
The Hadamard coin $H$ has the property that a leftward step followed by 
another leftward step produces a factor of $-1$ relative to other two-step 
sequences. This phase factor is the origin of quantum interference in the 
walk and is ultimately responsible for the ballistic spreading and the 
deviation from classical diffusive behavior.

Now we take the initial state $|\phi_\text{in}\ra = |R\ra \otimes |0\ra$ and explicitly
evalute the probability amplitude of $U^n |\phi_\text{in}\ra$ on the position :
\be
U^n |\phi_\text{in}\ra=\left(\frac{1}{\sqrt{2}}\right)^n\Big\{\binom{n}{0}|R\ra\otimes|n\ra+
\sum^n_{k=1}[a(n,k)|R\ra+b(n,k)|L\ra]\otimes|n-2k\ra\Big\}, \label{Uab}
\en
where 
 \be
a(n,k)&=&\sum^{k-1}_{l=0} (-1)^l\binom{k-1}{l}\binom{n-k}{k-l},\label{adefine}\\
b(n,k)&=&\sum^{k-1}_{l=0} (-1)^l \binom{k-1}{l} \binom{n-k}{k-l-1}, \label{bdefine}
 \en
and the convention $\binom{p}{q}=0$ for $q<0$ or $q>p$ is adopted. 
In Eq. (\ref{Uab}), 
the coefficient $\binom{n}{0}$ gives the contribution to the probability 
amplitude from the unique path consisting entirely of rightward steps; $k$ 
denotes the number of steps to the left, so that $n-k$ 
is the number of steps to the right, and $|n-2k\ra$ is the position state 
reached after taking $n$ steps; $a(n,k)$ and $b(n,k)$ correspond to the 
probability amplitudes for paths whose final step is rightward and leftward, 
respectively. This decomposition reflects the role of the coin state as an 
internal memory of the previous step's direction.  A closely related 
closed-form expression for the amplitudes appears in Lemma~4 of 
Ambainis et al.~\cite{ambainis2001}; our formulation is equivalent but is 
presented in a form that makes the underlying combinatorial structure 
explicit, which will be essential for the developments that follow. We note 
that the coefficients $(-1)^l \binom{k-1}{l}$ are precisely the $l$-th entry of 
the $(k-1)$-th row of the inverse Pascal triangle, reflecting the inclusion-
exclusion principle at work in the merging of leftward steps.

In more detail, for $a(n,k)$, when $l=0$ in Eq. (\ref{adefine}), the binomial 
coefficient $\binom{n-k}{k}$ means the following combinatorial picture: a 
sequence of $n-k$ rightward steps creates $n-k+1$ intervals. Excluding 
the rightmost interval (since the final step must be rightward), 
we choose $k$ of the remaining $n-k$ intervals and place one leftward 
step in each. The factor $(-1)^0$ reflects that no two leftward steps are 
adjacent (no $LL$ pairs) in this class of configurations, and the binomial 
coefficient $\binom{k-1}{0}=1$ indicates that none of the $k-1$ junctions 
between leftward steps are selected for merging. Similarly,  when $l=1$ in 
Eq.~(\ref{adefine}), the binomial coefficient $\binom{n-k}{k-1}$ means we 
choose $k-1$ of the $n-k$ intervals: in one of them we place a pair of 
consecutive leftward steps $LL$, and in each of the remaining $k-2$ 
intervals we place a single $L$. The factor $\binom{k-1}{1}=k-1$ counts 
the number of ways to choose which of the $k-1$ intervals receives the pair 
$LL$, and the prefactor $(-1)^1$ supplies the minus sign associated with 
each such configuration. Likewise, for $l=2$ in Eq. (\ref{adefine}), the 
binomial coefficient $\binom{n-k}{k-2}$ means we choose 
$k-2$ of the $n-k$ intervals.  Two distinct types of configurations 
contribute to this case:
\begin{itemize}
\item[(i)] Two of the $k-2$ intervals each receive a pair $LL$, and the 
remaining $k-4$ intervals each receive a single $L$. There are 
$\binom{k-2}{2}$ ways to choose which two intervals contain the $LL$ pairs.
\item[(ii)] One of the $k-2$ intervals receives a triple $LLL$, and the 
remaining $k-3$ intervals each receive a single $L$. There are 
$\binom{k-2}{1}$ ways to choose which interval contains the $LLL$.
\end{itemize}
Both types of configuration carry the same sign $(-1)^2=+1$. The total number 
of configurations is therefore $\binom{k-2}{2} + \binom{k-2}{1}$, which by 
the combinatorial identity $\binom{k-2}{2} + \binom{k-2}{1} = \binom{k-1}{2}$ 
is precisely the coefficient appearing in the amplitude. The overall 
contribution of this part is $(-1)^2 \binom{k-1}{2}$. The cases $l \ge 3$ are treated 
analogously, continuing until $l = k-1$. Whenever $n-k < k-l$, the corresponding 
term vanishes due to the convention $\binom{n-k}{k-l} = 0$.

Regarding $b(n,k)$, one leftward step already occupies the rightmost position. 
In Eq.~(\ref{bdefine}), when $l=0$, we select $k-1$ of the $n-k$ intervals 
and place one $L$ in each; the coefficient remains $(-1)^0 \binom{k-1}{0}=1$.
For $l=1$, we choose $k-2$ of the $n-k$ intervals to accommodate one $LL$ 
pair and the remaining single $L$'s. Importantly, the rightmost interval, 
which already contains a single $L$, can now be upgraded to host the $LL$ pair. 
This provides one additional choice, so that the total number of configurations 
is $\binom{k-1}{1}$, and the coefficient is $(-1)^1 \binom{k-1}{1}$.

In short, Eqs. (\ref{Uab})$\sim$(\ref{bdefine}) encode the complete information 
about the probability amplitudes of the quantum random walk after $n$ steps. 
It is worth noting that the expressions for $a(n,k)$ and $b(n,k)$ admit a natural 
interpretation in terms of discrete convolutions. Specifically, $a(n,k)$ can be 
written as the convolution of two sequences,
\begin{equation}
a(n,k) = \sum_{l=0}^{k-1} f_k(l) \, g_{n,k}(k-l), \qquad
f_k(l) = (-1)^l \binom{k-1}{l}, \quad
g_{n,k}(m) = \binom{n-k}{m},
\end{equation}
and similarly for $b(n,k)$ with $g_{n,k}(k-l-1)$. The sequence 
$f_k(l)$ carries the alternating signs arising from the merging of 
leftward steps and plays the role of a phase factor in the path 
integral, while $g_{n,k}(m)$ counts the number of unrestricted ways 
to distribute the remaining leftward steps. In this sense, the 
amplitudes $a$ and $b$ realize a discrete analogue of the Feynman 
path integral: a coherent sum over all possible histories, with 
interfering phases determined by the combinatorial structure of 
the paths. We note that a path-integral perspective on quantum walks 
was already implicit in the work of Meyer~\cite{meyer1996} on quantum 
lattice gases.

To illustrate Eq.~(\ref{Uab}), we explicitly evaluate it for $n=1,2,3$ 
in its unsimplified form:
\be
U |\phi_\text{in}\ra&=&\frac{1}{\sqrt{2}}\Big\{\binom{1}{0}|R\ra\otimes|1\ra+
\left[\binom{0}{1}|R\ra+\binom{0}{0}|L\ra\right]\otimes|-1\ra\Big\},\label{n=1} \\
U^2 |\phi_\text{in}\ra&=&\left(\frac{1}{\sqrt{2}}\right)^2 \Big\{\binom{2}{0}|R\ra\otimes|2\ra+
\left[\binom{1}{1}|R\ra+\binom{1}{0}|L\ra\right]\otimes|0\ra\non \\
 &&\quad\quad\quad +\left[\left(\binom{0}{2}-\binom{0}{1}\right)|R\ra+\left(\binom{0}{1}
-\binom{0}{0}\right)|L\ra\right]\otimes|-2\ra\Big\}, \label{n=2} \\
U^3 |\phi_\text{in}\ra&=&\left(\frac{1}{\sqrt{2}}\right)^3 \Big\{\binom{3}{0}|R\ra\otimes|3\ra+
\left[\binom{2}{1}|R\ra+\binom{2}{0}|L\ra\right]\otimes|1\ra\non \\
 &&\quad\quad\quad +\left[\left(\binom{1}{2}-\binom{1}{1}\right)|R\ra+\left(\binom{1}{1}
-\binom{1}{0}\right)|L\ra\right]\otimes|-1\ra\non \\
 &&\quad\quad\quad + \left[\left(\binom{0}{3}-2\binom{0}{2}+\binom{0}{1}\right)|R\ra
+\left(\binom{0}{2}-2\binom{0}{1}+\binom{0}{0}\right)|L\ra\right]\otimes|-3\ra\Big\}.\non\\ 
\label{n=3} 
\en
If we discard the alternating signs $(-1)^l$ and ignore the distinction 
between the coin states $|R\rangle$ and $|L\rangle$ in Eqs.~(\ref{n=1})
--(\ref{n=3})---thereby 
eliminating quantum interference and quantum memory---the coefficients 
at each position combine to give the classical binomial distribution:
\be
&&U |\phi_\text{in}\ra \to U |\phi_\text{in}\ra_{\text{cl}}=\frac{1}{\sqrt{2}}
\Big\{\binom{1}{0}|1\ra+\binom{0}{0}|-1\ra\Big\},\label{n=1c} \\
&&U^2 |\phi_\text{in}\ra \to U^2 |\phi_\text{in}\ra_{\text{cl}}=
\left(\frac{1}{\sqrt{2}}\right)^2 \Big\{\binom{2}{0}|2\ra+
\left[\binom{1}{1}+\binom{1}{0}\right]|0\ra +\binom{0}{0}|-2\ra\Big\}, 
\label{n=2c} \\
&&U^3 |\phi_\text{in}\ra \to U^3 |\phi_\text{in}\ra_{\text{cl}} \non \\
&&=\left(\frac{1}{\sqrt{2}}\right)^3 \Big\{\binom{3}{0}|3\ra+
\left[\binom{2}{1}+\binom{2}{0}\right]|1\ra +\left[\binom{1}{1}
+\binom{1}{1}+\binom{1}{0}\right]|-1\ra +\binom{0}{0}|-3\ra\Big\}.
\non\\ \label{n=3c} 
\en
Squaring the prefactor $(1/\sqrt{2})^{\,n}$ then yields the classical 
probability distribution $P_{\text{cl}}(n,x)$:
\be
P_{\text{cl}}(1,1)&=& \frac{1}{2},~~P_{\text{cl}}(1,-1) = \frac{1}{2}. \label{Pcl1}\\
P_{\text{cl}}(2,2)&=& \frac{1}{4},~~P_{\text{cl}}(2,0) = \frac{2}{4},~~P_{\text{cl}}(2,-2) 
= \frac{1}{4}. \label{Pcl2}\\
P_{\text{cl}}(3,3) &=& \frac{1}{8},~~P_{\text{cl}}(3,1) = \frac{3}{8},~~P_{\text{cl}}(3,-1) 
= \frac{3}{8}, ~~P_{\text{cl}}(3,-3) = \frac{1}{8}.\label{Pcl3}
\en
These mimic the continuous measurement processes those destroy quantum 
coherence at each step. In contrast, retaining the $(-1)^l$ factors and the coin-
state distinction preserves quantum coherence, the simplified equations of 
Eqs.~(\ref{n=1})--(\ref{n=3}) are
\be
U |\phi_\text{in}\ra&=&\frac{1}{\sqrt{2}}\Big\{\binom{1}{0}|R\ra\otimes|1\ra
+\binom{0}{0}|L\ra\otimes|-1\ra\Big\},\label{n=1q} \\
U^2 |\phi_\text{in}\ra&=&\left(\frac{1}{\sqrt{2}}\right)^2 \Big\{\binom{2}{0}
|R\ra\otimes|2\ra+
\left[\binom{1}{1}|R\ra+\binom{1}{0}|L\ra\right]\otimes|0\ra-\binom{0}{0}
|L\ra\otimes|-2\ra\Big\}, \quad\quad \label{n=2q} \\
U^3 |\phi_\text{in}\ra&=&\left(\frac{1}{\sqrt{2}}\right)^3 \Big\{\binom{3}{0}
|R\ra\otimes|3\ra+
\left[\binom{2}{1}|R\ra+\binom{2}{0}|L\ra\right]\otimes|1\ra-\binom{1}{1}
|R\ra\otimes|-1\ra\non \\
 &&\quad\quad\quad\quad+ \binom{0}{0}|L\ra\otimes|-3\ra\Big\}. \label{n=3q} 
\en
The quantum probability distribution $P_{\text{q}}(n,x)$ is obtained by squaring the 
amplitudes of the two coin states at each position and summing them:
\be
P_{\text{q}}(1,1)&=& \frac{1}{2},~~P_{\text{q}}(1,-1) = \frac{1}{2}.\label{Pqk1}\\
P_{\text{q}}(2,2)&=& \frac{1}{4},~~P_{\text{q}}(2,0) = \frac{2}{4},~~P_{\text{q}}(2,-2) 
= \frac{1}{4}. \label{Pq2}\\
P_{\text{q}}(3,3) &=& \frac{1}{8},~~P_{\text{q}}(3,1) = \frac{5}{8},~~P_{\text{q}}(3,-1) 
= \frac{1}{8}, ~~P_{\text{q}}(3,-3) = \frac{1}{8}. \label{Pq3}
\en
The above calculations reveal why the QRW and the RW agree 
for $n \le 2$. For $n=2$, a pair of consecutive leftward steps does occur and 
does contribute a minus sign via the factor $(-1)^l$. However, only a single 
path reaches the position $x=-2$, so there is no other path to interfere with; 
the minus sign disappears upon squaring, and the resulting 
probability coincides with the classical value. The deviation at $n=3$ arises 
because multiple interfering paths now contribute to the same final position, 
and the relative signs between them lead to constructive or destructive 
interference.

To understand the general case, we now verify that
discarding the alternating factors $(-1)^l$ from the QRW amplitudes recovers the 
classical binomial distribution for all $n$. Concretely, let $\tilde{a}(n,k)$ 
and $\tilde{b}(n,k)$ denote the expressions obtained from Eqs.~(\ref{adefine}) 
and (\ref{bdefine}) by discarding those signs. Summing over the two coin 
states then gives
\be
\tilde{a}(n,k) + \tilde{b}(n,k) &=& \sum^{k-1}_{l=0} \binom{k-1}{l}
\left[\binom{n-k}{k-l}+\binom{n-k}{k-l-1}\right]\non \\
&=&\sum^{k}_{l=0} \binom{k-1}{l}\binom{n-k+1}{k-l}=\binom{n}{k},
\en
where Vandermonde's convolution is used. The sum $\tilde{a}+\tilde{b} 
= \binom{n}{k}$, the number of possible paths, when multiplied by the 
squared normalization factor $(1/\sqrt{2})^{2n}$, reproduces the classical 
probability $P_{\text{cl}}(n,x) = \binom{n}{k}/2^n$, 
with $x = n-2k$. This confirms that the classical walk emerges from the 
quantum walk when the phase factors responsible for interference are erased.


\section{From Krawtchouk Matrices to Parseval Identity}
For the RW, the total probability is conserved:
\be
\sum^n_{x=-n} P_\text{cl}(n,x)=\sum_{k=0}^n \frac{1}{2^n} \binom{n}{k} = 1,
\en
as follows immediately from the binomial theorem. We now show that the 
quantum counterpart satisfies an analogous but non-trivial conservation 
law---a Parseval identity, 
\be
\sum^n_{x=-n} P_\text{q}(n,x)=\frac{1}{2^n}\Big\{1+\sum_{k=1}^{n}
 \big[ a(n,k)^2 + b(n,k)^2 \big] \Big\}= 1,
\label{Pq1}
\en
connecting the amplitudes $a(n,k)$ and $b(n,k)$ to the Krawtchouk matrix. 
Although the QRW is built from manifestly unitary coin and shift operators, 
it is not a priori obvious that this unitarity survives when the amplitudes are 
expressed through the Krawtchouk matrix elements; the Parseval 
identity established in this section confirms that it does.

To this end, we introduce the generating function for the element of 
Krawtchouk matrix $K^{(n)}_{ij}$: 
\be
(1-z)^j(1+z)^{n-j}=\sum^n_{i=0} K^{(n)}_{ij}z^i ,\label{Kgen}
\en
where the indices $i$ (row) and $j$ (column) run from $0$ to $n$. Expanding the 
left-hand side of Eq. (\ref{Kgen}) yields the expression
\be
K^{(n)}_{ij}=[z^i](1-z)^j(1+z)^{n-j}=\sum_{l=0}^{\text{min}(i,j)} (-1)^l 
\binom{j}{l}\binom{n-j}{i-l}. \label{Kdef}
\en
For example, when $n=2,3$, the Krawtchouk matrices are
\be
K^{(2)} = \left[\begin{array}{rrr}
                             1 &  1 &  1 \cr
                             2 &  0 & -2 \cr
                             1 & -1 &  1 \cr \end{array}\right],~~\label{n2}
K^{(3)} = \left[\begin{array}{rrrr}
                             1 &  1 &  1 & 1 \cr
                             3 &  1 & -1 & -3 \cr
                             3 & -1 & -1 & 3 \cr
                             1 & -1 & 1 & -1 \cr \end{array}\right].\label{n3}
\en
The matrix elements $K^{(n)}_{ij}$ are precisely the symmetric Krawtchouk 
polynomials $\kappa_i(j; n)$ evaluated at the integer lattice points, i.e., 
$K^{(n)}_{ij} = \kappa_i(j; n)$. From the definition, the entries of the Krawtchouk 
matrix satisfy the following cross identities \cite{feinsilver2007}, which will be 
used in the proof of the Parseval identity below:
\be
&(i)& K_{i-1,j}^{(n)}+ K_{i,j}^{(n)} = K_{i,j}^{(n+1)},~~~~~ 
(ii)~~ K_{i,j}^{(n)}+ K_{i,j+1}^{(n)} =2 K_{i,j}^{(n-1)}, \non\\  
&(iii)& K_{i,j}^{(n)} - K_{i-1,j}^{(n)} = K_{i,j+1}^{(n+1)},~~~~~
(iv)~~ K_{i,j}^{(n)} - K_{i,j+1}^{(n)} =2 K_{i-1,j}^{(n-1)}. \label{crossid}
\en
\def\b{\circ}   \def\a{\bullet}
\newcommand{\bigsquare}{\,\fbox{\rule{0em}{0.35em}\rule{0.4em}{0em}}\,}
\newcommand{\squareacircle}{\,\framebox[1.0em]{\makebox[0em]{$\a$}}\,}
\newcommand{\squarecircle}{\,\framebox[1.0em]{\makebox[0em]{$\b$}}\,}
The relation of this matrix to the QRW amplitudes becomes clear when we 
compare the generating functions of $a$ and $b$ with that of $K^{(n)}_{ij}$:
\be
a(n,k)+b(n,k) &=& K_{k, k-1}^{(n)},\label{abplus}\\
a(n,k)-b(n,k) &=& K_{k, k}^{(n)}.\label{abminus}
\en
While Eq. (\ref{abplus}) can also be obtained directly from the combinatorial 
definitions, the antisymmetric relation Eq. (\ref{abminus}) relies on the generating 
function. The detailed derivation is given in Appendix~\ref{app:genfun}. We note, 
however, that a simpler derivation of Eq. (\ref{abminus}) emerges once the 
difference operator $D_k$ is introduced in Section IV. In addition, Eqs. (\ref{abplus})
and (\ref{abminus}) reveal that the symmetric and antisymmetric combinations of 
the QRW amplitudes correspond precisely to the diagonal ($\a$) and the adjacent 
left-diagonal ($\b$) elements of $K^{(n)}$, while the classical binomial distribution 
uses only the leftmost column ($ \bigsquare$) $K^{(n)}_{k,0} = \binom{n}{k}$:
\be
K^{(n)} = \left[\begin{array}{cccc}
                        ~~ \squareacircle ~ ~  &   ~~  &   ~~  &~~ \cr
                      ~~   \squarecircle~~&~~ \a ~~&   ~~ &  ~~  \cr
                      ~~    \bigsquare ~~    &~~ \b~~&~~\a  ~~&~~ \cr 
                      ~~     \vdots         ~~          & ~~  &~\ddots~~ &~\ddots  ~~\cr
\end{array}\right].\non
\en
Solving the simultaneous equations Eqs. (\ref{abplus})
and (\ref{abminus}), we obtain
\be
a(n,k) = K^{(n-1)}_{k, k-1}, \qquad b(n,k) = K^{(n-1)}_{k-1, k-1}. \label{abK}
\en
These relations connect the Krawtchouk matrices at successive time 
steps. To illustrate this connection, we display the two matrices $K^{(n-1)}$ 
and $K^{(n)}$ for $k=2$:
\be
K^{(n-1)} = \left[\begin{array}{cccc}
                      ~~ \squareacircle~~  &   ~~  &   ~~   & ~~  \cr
                       ~~   \squarecircle  ~~&~~b ~~&   ~~ &  ~~   \cr
                      ~~    \bigsquare ~~    &~~ a~~&~~\a  ~~&~~ \cr
                      ~~     \vdots         ~~          & ~~   &~\ddots~~ &~\ddots  ~~\cr
\end{array}\right]
\quad\longrightarrow\quad
K^{(n)} = \left[\begin{array}{cccc}
                        ~~ \squareacircle ~ ~  &   ~~  &   ~~   & ~~  \cr
                      ~~   \squarecircle~~&~~ \a ~~&   ~~ &  ~~   \cr
                      ~~    \bigsquare ~~    &a+b&~a-b  ~&~~ \cr
                           \vdots                  &   &\ddots~~~~~ &\ddots~~~~~  \cr
\end{array}\right].\non
\en
The amplitudes $a(n,k)$ and $b(n,k)$ at step $n$ are found in the diagonal 
and sub-diagonal of $K^{(n-1)}$, while their sum and difference occupy the 
corresponding positions in $K^{(n)}$. This shift from $K^{(n-1)}$ to $K^{(n)}$ 
is the key ingredient that makes the inductive proof of the Parseval identity 
possible. We now proceed to that proof.
From the Eqs. (\ref{abplus}) and (\ref{abminus}), our aim is to prove the Parseval
identity
\be
\sum^{n}_{k=1} \frac{1}{2} \left[\left(K^{(n)}_{k,k-1}\right)^2+
\left(K^{(n)}_{k,k}\right)^2\right]=2^n-1. 
\label{aim}
\en
We proceed by induction on $n$. For the case $n=3$, from $(ii)$ and $(iv)$ of the 
Eq. (\ref{crossid}), Eq. (\ref{aim}) gives the result
\be
&&\sum^{3}_{k=1} \frac{1}{2} \left[\left(K^{(3)}_{k,k-1}\right)^2
+\left(K^{(3)}_{k,k}\right)^2\right]
=\sum^{3}_{k=1}  \left[\left(\frac{K^{(3)}_{k,k-1}+K^{(3)}_{k,k}}{2}\right)^2
+\left(\frac{K^{(3)}_{k,k-1}-K^{(3)}_{k,k}}{2}\right)^2\right]\non \\
&=&\left[\left(K^{(2)}_{1,0}\right)^2+\left(K^{(2)}_{0,0}\right)^2+\left(K^{(2)}_{2,1}\right)^2
+\left(K^{(2)}_{1,1}\right)^2+\left(K^{(2)}_{2,2}\right)^2\right].
\en
which evaluates to $2^3 - 1 $ using the explicit entries of $K^{(2)}$ (Eq. (\ref{n2})).
Assume that Eq. (\ref{aim}) holds for $n = m$, i.e.,
\be
\sum^{m}_{k=1} \frac{1}{2} \left[\left(K^{(m)}_{k,k-1}\right)^2
+\left(K^{(m)}_{k,k}\right)^2\right]=2^m-1. 
\label{mp0}
\en
We now show that it also holds for $n = m+1$:
\be
&&\!\!\!\!\sum^{m+1}_{k=1} \frac{1}{2} \left[\left(K^{(m+1)}_{k,k-1}\right)^2
+\left(K^{(m+1)}_{k,k}\right)^2\right]
=\sum^{m+1}_{k=1}  \left[\left(\frac{K^{(m+1)}_{k,k-1}
+K^{(m+1)}_{k,k}}{2}\right)^2
+\left(\frac{K^{(m+1)}_{k,k-1}-K^{(m+1)}_{k,k}}{2}\right)^2\right]\non \\
&&=\Big[\left(K^{(m)}_{1,0}\right)^2+\left(K^{(m)}_{0,0}\right)^2
+\left(K^{(m)}_{2,1}\right)^2
+\left(K^{(m)}_{1,1}\right)^2+\left(K^{(m)}_{3,2}\right)^2\left(K^{(m)}_{2,2}\right)^2+...
\non \\
&&+\left(K^{(m)}_{m-1,m-2}\right)^2+\left(K^{(m)}_{m-2,m-2}\right)^2
+\left(K^{(m)}_{m,m-1}\right)^2
+\left(K^{(m)}_{m-1,m-1}\right)^2+\left(K^{(m)}_{m,m}\right)^2\Big],\label{mp1}
\en
where $K^{(m)}_{m+1,m}=0$. At this point, we isolate the term 
$\big(K^{(m)}_{0,0}\big)^2$ in Eq.~(\ref{mp1}) and recombine the remaining 
terms in pairs (e.g., the 1st with the 4th, the 3rd with 
the 6th, and so on). This pairing yields a sum over $k' = 1\sim m$, then 
\be
\sum^{m+1}_{k=1} \frac{1}{2} \left[\left(K^{(m+1)}_{k,k-1}\right)^2
+\left(K^{(m+1)}_{k,k}\right)^2\right]
=\left( K^{(m)}_{0,0}\right)^2+\sum^{m}_{k'=1}\left[\left(K^{(m)}_{k',k'-1}\right)^2+
\left(K^{(m)}_{k',k'}\right)^2\right]\label{recom}
\en
Compare Eq. (\ref{mp0}) with the second term of the right-hand side of 
Eq. (\ref{recom}), we can obtain
\be
\sum^{m+1}_{k=1} \frac{1}{2} \left[\left(K^{(m+1)}_{k,k-1}\right)^2
+\left(K^{(m+1)}_{k,k}\right)^2\right]
=1+2\times (2^m-1)=2^{m+1}-1.
\en
Thus, by induction, Eq.~(\ref{aim}) holds for all $n \ge 3$. Together with 
Eqs.~(\ref{abplus}) and (\ref{abminus}), 
this yields the Parseval identity for the QRW amplitudes, Eq. (\ref{Pq1}).


\section{From Difference operator to Hermite limits}
The coefficients $(-1)^l \binom{k-1}{l}$ appearing in the definitions 
of $a(n,k)$ and $b(n,k)$, i.e., Eqs. (\ref{adefine}) and (\ref{bdefine}), are the 
entries of the $(k-1)$-th row of the inverse Pascal triangle. This structure 
is characteristic of backward differences acting on binomial coefficients, 
and it motivates the introduction of a modified difference operator $D_k$ 
tailored to the combinatorial structure of the QRW amplitudes. Accordingly,
we define 
\be
D_k \binom{n-k}{k}\equiv  \binom{n-k}{k}-\binom{n-k}{k-1},\label{Ddef}
\en
where the definition applies for $0 \le k \le n/2$; for $k > n/2$, the 
binomial coefficient $\binom{n-k}{k}$ vanishes identically.
$D_k$ is different from the traditional backward difference operator $\nabla_k$ 
in that $D_k$ acts with $N = n-k$ held fixed. Equivalently, $D_k$ and $\nabla_k$ 
coincide when applied to $\binom{n}{k}$. From the definition of $D_k$,
we immediately obtain
\be
a(n,k)&=&D^{k-1}_k \binom{n-k}{k}= K_{k, k-1}^{(n-1)}, \label{an1}\\
b(n,k)&=&D^{k-1}_k \binom{n-k}{k-1}=K_{k-1, k-1}^{(n-1)}. \label{bn1}
\en
A simple manipulation then yields the symmetric and antisymmetric 
combinations
\be
a(n,k)+b(n,k)&=& D^{k-1}_k\Big[ \binom{n-k}{k}+\binom{n-k}{k-1}\Big]\non \\
&=& D^{k-1}_k\binom{n-k+1}{k}=K^{(n)}_{k,k-1}, \label{apb}\\
a(n,k)-b(n,k)&=& D^{k-1}_k\Big[ \binom{n-k}{k}-\binom{n-k}{k-1}\Big]\non \\
&=& D^{k-1}_kD_k \binom{n-k}{k}=D^{k}_k\binom{n-k}{k}=K^{(n)}_{k,k}.\label{amb}
\en
These relations, including the last one for $a-b$, now follow without 
recourse to the generating function used in Section~III.
Up to this point, the operator $D_k$ may appear to serve only as a 
bookkeeping tool for linking the amplitudes $a$ and $b$ to the entries 
of the Krawtchouk matrix. In fact, however, it plays several further 
roles that are essential to the structure of the QRW. 

As noted earlier, the RW distribution involves only the 
leftmost column of the Krawtchouk matrix, namely $K^{(n)}_{k,0} = 
\binom{n}{k}$. We begin by recalling the well-known continuous limit 
of the RW. For large $n$, the binomial distribution 
approaches a Gaussian,
\be
\frac{1}{2^n} \binom{n}{\frac{n-x}{2}} \xrightarrow{n \to \infty} 
\frac{1}{\sqrt{2\pi n}} \exp\!\Big( -\frac{x^2}{2n} \Big),
\en
with $x = n-2k$ and the usual $\sqrt{n}$ scaling. It is convenient 
to introduce the rescaled variable $\xi = \frac{x}{\sqrt{2n}}$,
so that the Gaussian weight takes the standard form $e^{-\xi^2}$. 
The Hermite polynomials $H_m(\xi)$ are then generated by the Rodrigues 
formula
\be
H_m(\xi) = (-1)^m e^{\xi^2} \frac{d^m}{d\xi^m} e^{-\xi^2}.
\en

We now show that the same structure emerges from the discrete difference 
operator $D_k$. Recall from the definition of $D_k$, the quantities
\be
h_1(n,k) = \binom{n}{k}^{-1} D_k{\binom{n}{k}} = \frac{n-2k+1}{n-k+1}, \label{h1}
\en
and
\be
h_2(n,k) = \binom{n}{k}^{-1} D_k^2{\binom{n}{k}} = 
\frac{(n-2k+1)(n-2k+2)-2k}{(n-k+1)(n-k+2)},
\en
which measure the relative change produced by single and double applications
of $D_k$ on the binomial coefficient, respectively. $\binom{n}{k}^{-1}$ denotes the 
multiplicative inverse (not the functional inverse). This notation is chosen 
to facilitate comparison with the continuous case, where the Hermite 
polynomials are generated by the operator $e^{\xi^2} \frac{d}{d\xi} 
$ acting on the Gaussian weight $e^{-\xi^2}$, and the factor 
$e^{\xi^2}$ plays the role of the reciprocal of the weight.
In the limit $n \to \infty$ with $\xi = x/\sqrt{2n}$ held fixed, 
a direct calculation yields
\be
h_1(n,k)& \xrightarrow{n \to \infty}& \sqrt{\frac{2}{n}} H_1(\xi), \\
h_2(n,k)& \xrightarrow{n \to \infty} &\left(\sqrt{\frac{2}{n}} \right)^2H_2(\xi),
\en
where $H_1(\xi) = 2\xi$ and $H_2(\xi)=4\xi^2-2$ are the first and second 
Hermite polynomials, respectively. The operator $D_k$ satisfies a discrete 
analogue of the Leibniz rule: for any functions $f$ and $g$,
\be
D_k(f \cdot g) = (D_k f) \cdot g + f \cdot (D_k g) - (D_k f) \cdot (D_k g).
\en
The additional term $- (D_k f) \cdot (D_k g)$ is characteristic of 
discrete difference operators and vanishes in the continuous limit, 
where the usual product rule is recovered. This identity underpins 
the recurrence relations among the $h_m$ and reinforces the 
interpretation of $D_k$ as a genuine discrete derivative.
More generally, for the $m$-th iterated difference,
\be
h_m(n,k) &=&\binom{n}{k}^{-1} D_k^m \binom{n}{k}\non \\
&=& \frac{(n-k)!}{(n-k+m)!} 
\sum_{j=0}^{\lfloor m/2 \rfloor} 
(-1)^j \frac{m!}{(m-2j)! \, j!} 
\, \frac{k!}{(k-j)!} 
\, \frac{(n-2k+m)!}{(n-2k+2j)!} \label{hm exact} \\
&\xrightarrow{n \to \infty} &
\left( \sqrt{\frac{2}{n}}  \right)^m H_m(\xi),\non
\en
where 
\be
H_m(\xi)= \sum_{j=0}^{\lfloor m/2 \rfloor} 
(-1)^j \frac{m!}{(m-2j)!\, j!}\, (2\xi)^{m-2j}.
\en
The structural parallel between $h_m$ and $H_m$ is evident: the 
coefficients coincide, and the powers $(2\xi)^{m-2j}$ are replaced 
in the discrete case by ratios of factorials that reduce to the 
same powers in the continuous limit.
Thus the operator $D_k$ acts as a discrete analogue of the Hermite 
polynomial generator, with the continuous limit recovering the 
standard Rodrigues formula. For general $m$, the normalization condition 
for $h_m$ reads
\begin{equation}
\frac{1}{2^n} \sum_{k=0}^{n} \frac{(n-k+m)!}{2^m(n-k)!}
\, h_m(n,k)^2
\, \binom{n}{k} \approx 1,
\label{eq:norm_hm}
\end{equation}
where the factor $2^m$ in the denominator compensates for a systematic 
half-integer shift of $m/2$ steps originating from the numerator 
$(n-2k+1)\cdots(n-2k+m)$ of $h_m$, which is centered at $x + m/2$ 
rather than $x$. The slight deviation of the normalization from unity for finite 
$n$ is thus a discretization artifact that vanishes in the 
continuous limit, where $2^{-m}(n-k+1)^{\overline{m}} \sim (n/2)^m$ 
combines with $h_m \sim (\sqrt{2/n})^m H_m$ to reproduce the 
standard Hermite normalization.

This result aligns naturally with the Askey limit of the Krawtchouk 
polynomials. Under the same scaling, the symmetric Krawtchouk 
polynomials $\kappa_k(j; n)$ converge to the Hermite polynomials
, a classic result in the Askey scheme  \cite{askey1985, koekoek2010}:
\be
\lim_{n \to \infty} \sqrt{\binom{n}{k}} \, \kappa_k\!\left( \frac{n}{2} +
 \frac{\xi}{2}\sqrt{n}; n \right) = \frac{1}{\sqrt{2^k k!}} H_k(\xi).
\en
Since $a(n,k)$ and $b(n,k)$ are linear combinations of $\kappa_k(k-1; n)$ 
and $\kappa_k(k; n)$, the convergence of $h_m$ to $H_m$ can be viewed 
as a direct consequence of the Askey limit, providing an independent 
check on our discrete formulation. 

We note that the quantities $h_m(n,k)$ are defined with respect to 
$\binom{n}{k}$, whereas the QRW amplitudes involve $\binom{n-k}{k}$. 
The two are isomorphic under the substitution $n \mapsto n-k$, that is, 
replacing the total number of steps $n$ by $N = n-k$.\footnote{For example, 
$h_1(n-k,k) = \frac{n-3k+1}{n-2k+1}$ is obtained 
from $h_1(n,k) = \frac{n-2k+1}{n-k+1}$ by the substitution $n \mapsto n-k$.}
Since 
the the structure of $D_k$ depends only on $N$ and $k$, all results 
obtained for $h_m(n,k)$ carry over directly to $a(n,k)$ and $b(n,k)$.

Having established the continuous limit of 
$D_k$, we now return to the discrete setting to uncover its role as a raising 
operator on the Krawtchouk matrices.


\section{From RW to QRW? $D_k$ as a Coherence Generator }
In the previous section, we introduced the difference operator $D_k$ 
and established its continuous limit. We now reveal a deeper role of 
$D_k$ at the discrete level: it connects successive Krawtchouk matrices 
and acts as a raising operator. From  Eq. (\ref{abK}),
$a(n,k) $ and $ b(n,k)$ are again the diagonal and adjacent left-diagonal 
entries, but of the Krawtchouk matrix at step $n-1$. That is, the QRW 
amplitudes after $n$ steps are in fact determined by the Krawtchouk 
matrix of the previous step $n-1$. This observation not only reflects the 
memory property of the QRW coin states, but also highlights a deeper 
role played by $D_k$---namely, connecting the Krawtchouk matrices at 
consecutive time steps. For convenience, we illustrate this with the example 
$n=6$. Table~\ref{tab:n6} shows how the operator $D_k$ generates the QRW 
amplitudes from the classical binomial distribution. The boxed entries 
are the RW amplitudes $\boxed{K^{(n)}_{k,0}} = \binom{n}{k}$; 
$\a$ and $\b$ denote the diagonal and adjacent left-diagonal 
elements, respectively. Each arrow ($\rightsquigarrow$) represents 
one application of $D_k$, which acts as a coherence generator, 
converting the number of possible paths in the classical walk into 
higher-order interfering amplitudes. The corresponding values of 
$h_m$ are shown in the rows below. In this sense, $D_k$ plays a role 
analogous to the Hadamard gate in circuit-based quantum 
computation: it transforms a classical binomial distribution into a 
quantum amplitude carrying phase information.
\begin{table}[ht]
\begin{ruledtabular}
{\scriptsize
\begin{tabular}{c|r|r|r}
 $k$ & $1(x=4)$ & $2(x=2)$ & $3(x=0)$  \\\hline
$a+b$ & $\raisebox{-3.0pt}{\boxed{K^{(6)}_{1,0}}}\b$   &  
$\raisebox{-3.0pt}{\boxed{K^{(5)}_{2,0}}}
~\rightsquigarrow ~K^{(6)}_{2,1}\b$  & 
$\raisebox{-3.0pt}{\boxed{K^{(4)}_{3,0}}}
~(\rightsquigarrow)^2 K^{(6)}_{3,2}\b$ \\
$h_m$ & $h_0(6,1)$ & $ h_1(5,2)$~~~~~~~~~~~~~ 
& $ h_2(4,3)$~~~~~~~~~~~~~~ \\\hline
$a-b$ & $\raisebox{-3.0pt}{\boxed{K^{(5)}_{1,0}}} 
\rightsquigarrow K^{(6)}_{1,1}\a$   & 
$\raisebox{-3.0pt}{\boxed{K^{(4)}_{2,0}}}~(\rightsquigarrow )^2
K^{(6)}_{2,2}\a$ & $\raisebox{-3.0pt}{\boxed{K^{(3)}_{3,0}}}
~(\rightsquigarrow )^3 K^{(6)}_{3,3}\a$ \\
$h_m$ & $h_1(5,1)$ ~~~~~~~~~~~& $ h_2(4,2)$~~~~~~~~~~~~~
 & $h_3(3,3)$~~~~~~~~~~~~~~~\\\hline\hline
 $k$ & $4(x=-2)$ & $5(x=-4)$ & $6(x=-6)$  \\\hline
$a+b$ & $\raisebox{-3.0pt}{\boxed{K^{(4)}_{4,0}}}-
\raisebox{-3.0pt}{\boxed{K^{(3)}_{3,0}}} ~(\rightsquigarrow)^3 K^{(6)}_{4,3}\b$  
 & $\sum_{j=0}^{3} (-1)^j \binom{3}{j}~  \raisebox{-3.0pt}{\boxed{K^{(5-j)}_{5-j, 0}}}
~(\rightsquigarrow)^4 K^{(6)}_{5,4}\b$ & $\sum_{j=0}^{5} (-1)^j \binom{5}{j}~ 
 \raisebox{-3.0pt}{\boxed{K^{(6-j)}_{6-j, 0}}}~
(\rightsquigarrow )^5 K^{(6)}_{6,5}\b$\\
$h_m$ & $h_3(4,4)$ ~$h_3(3,3)$~~~~~~~~~~~~~~~& $h_4(5-j,5-j)$
~~~~~~~~~~~ & $h_5(6-j,6-j)$~~~~~~~~~~~~\\\hline
$a-b$ &  $\raisebox{-3.0pt}{\boxed{K^{(4)}_{4,0}}}-2~
\raisebox{-3.0pt}{\boxed{K^{(3)}_{3,0}}}
+\raisebox{-3.0pt}{\boxed{K^{(2)}_{2,0}}} ~(\rightsquigarrow)^4 K^{(6)}_{4,4}\a$   
& $\sum_{j=0}^{4} (-1)^j \binom{4}{j}~  \raisebox{-3.0pt}{\boxed{K^{(5-j)}_{5-j, 0}}}~
(\rightsquigarrow )^5 K^{(6)}_{5,5}\a$ & $\sum_{j=0}^{6} (-1)^j \binom{6}{j}~  
\raisebox{-3.0pt}{\boxed{K^{(6-j)}_{6-j, 0}}}~
(\rightsquigarrow )^6 K^{(6)}_{6,6}\a$ \\
$h_m$ & $h_4(4,4)~~~~~ h_4(3,3)~~~ h_4(2,2)$~~~~~~~~~~~~ ~~& 
$h_5(5-j,5-j)$~~~~~~~~~~~ & $h_6(6-j,6-j)$~~~~~~~~~~~
\end{tabular}}
\end{ruledtabular}
\caption{ Action of $D_k$ on the Krawtchouk matrix elements for $n=6$. 
Multiplying  each $\boxed{\text{boxed}}$ element by 
the $h_m$ shown directly below yields the QRW amplitude on the 
rightmost side of the corresponding cell ($\b$ or $\a$).}
\label{tab:n6}
\end{table}
In the upper half of Table \ref{tab:n6} ($x \ge 0$), 
each QRW amplitude is generated by a single $h_m$ acting on a classical 
binomial coefficient. The situation is more involved in the lower half ($x<0$):
the rightmost Krawtchouk matrix element ($\b$ or $\a$) in each 
cell arise from a difference 
structure of the $h_m$, in contrast to the direct multiplicative 
form that governs the upper half. Specifically, the Krawtchouk matrix 
elements for $x < 0$ can be expressed as alternating sums of 
$h_m$-weighted binomial coefficients, e.g., 
\be
K^{(6)}_{4,3} &=& K^{(7)}_{4,3} - K^{(6)}_{3,3} = D^3_k \binom{4}{4}-
D^3_k\binom{3}{3} \non \\
&=& h_3(4,4)~K^{(4)}_{4,0} - h_3(3,3)~K^{(3)}_{3,0}.\label{manyhm}
\en
The appearance of $K^{(7)}_{4,3}$ in the expression for 
$K^{(6)}_{4,3}$ reflects the difference structure of the $h_m$ 
in the left half: the QRW amplitudes for $x < 0$ can be built from 
contributions that, individually, correspond to larger values of 
$n$. (This is reminiscent of the relation $a(n,k) = K^{(n-1)}_{k,k-1}$ 
and $b(n,k) = K^{(n-1)}_{k-1,k-1}$ encountered in Section~III, where 
the amplitudes at step $n$ are also expressed through the Krawtchouk 
matrix of the previous step.) This is not a violation of causality but a 
consequence of the unitary constraints that link successive time steps.
Once the initial state and the coin operator are fixed, the entire 
unitary evolution $U^n$ is determined, and with it the probability 
amplitudes at all future times. The probability distribution at 
any step $n$ is thus fully encoded in the initial condition and the 
structure of the walk.

In the  RW, increasing the number of steps $n$ merely 
broadens the Gaussian profile while keeping the distribution in its 
ground state. In contrast, the operator $D_k$ generates genuinely new 
states---the discrete analogues of Hermite excited states---at a fixed 
$n$. The index $m$ in $h_m$ labels the degree of quantum coherence: 
larger $m$ corresponds to more frequent merging of leftward steps, 
i.e., stronger quantum interference. Thus, $D_k$ does not simply evolve 
the walk forward in time; it excites the walk into higher coherent 
modes within the same time slice. 

A glance at the Table \ref{tab:n6} reveals that all Krawtchouk matrix 
elements relevant to the QRW probability amplitudes, without exception, can 
be obtained from the classical binomial coefficients and the 
corresponding $h_m$. For $x \ge 0$, each Krawtchouk matrix element can 
be generated from a unique classical binomial distribution at a smaller value 
of $n$. For $x < 0$, by contrast, the element requires a combination of 
several classical distributions, each with $n$ smaller than or equal 
to the original, weighted by the inverse Pascal triangle and then 
coherently superimposed. Conversely, starting from the classical binomial 
distribution, one can reach any QRW probability amplitude; the only 
difference is whether the destination is reached through a single $h_m$ or a 
coherent combination of several.

\begin{figure}
 \includegraphics*[width=4.5in]{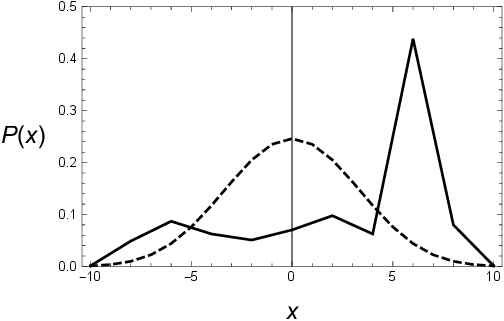}
 \caption{Probability distribution of the QRW (solid line) and the
RW (dashed line) for $n=10$.}
  \label{fig:1}
 \end{figure}

These alternating signs of Eq.~(\ref{manyhm}) lead to partial cancellations 
that smooth out the amplitudes, making the left ($x<0$) peak less 
sharp than its right ($x \ge 0$) counterpart. Fig.~\ref{fig:1} 
illustrates this distinction for $n=10$: the QRW distribution 
(solid line) displays two prominent ballistic peaks of unequal 
height and sharpness---the right peak being taller and narrower, 
the left peak broader and lower---whereas the RW 
(dashed line) remains a single Gaussian-like ground state.
The right peak arises because near the outermost positions $h_m$ is 
close to unity (e.g., Eq.~(\ref{h1}) gives $h_1 = 8/9$ for $n=10, k=1$) 
while the binomial coefficient is large. The left 
peak results from the inverse Pascal combination of several $h_m$, 
whose alternating signs shift the relative maximum away from the extreme 
left.

We now analyze how the quantities $h_m$, generated by the operator 
$D_k$, shape the ballistic spreading of the QRW. 
Although $h_m$ is now defined on both halves, the right half 
dominates the variance. As shown in Table~\ref{tab:right_all}, 
the right-half fraction of $\langle x^2 \rangle$ is $70.0\%$ at 
$n=6$ and $74.8\%$ at $n=10$, while the RW contributes 
equally from both halves. We therefore first focus our detailed analysis 
on $x \ge 0$, where each amplitude involves a single $h_m$, and 
then turn to the left half.
\begin{table}[h]
\centering
\begin{tabular}{c|c|c|c|c|c}
\hline\hline
$n$ &$\langle x^2 \rangle_{\text{QRW}}/n^2$ &$\langle x^2 \rangle_{\text{QRW}}$ (full) 
& $\langle x^2 \rangle_{\text{QRW}, x\ge 0}$ &
$x\ge 0$/full (QRW) & $\langle x^2 \rangle_{\text{RW}, x\ge 0}$ \\
\hline
6  & 0.313 & 11.25 & 7.88 & 70.0\%  & 3.00  \\
10 & 0.300 &29.95 & 22.40 & 74.8\% & 5.00 \\ 
$\infty$ & 0.293 & $(1-1/\sqrt{2}) n^2$ & — & — & $n/2$\\
\hline\hline
\end{tabular}
\caption{Comparison of $\langle x^2 \rangle$ for the QRW and the 
RW. The column $\langle x^2 \rangle_{x\ge 0}$ uses only the right half of 
the distribution but is normalized by the total probability (QRW) or by 
half the total probability (RW). The asymptotic value for $n\to\infty$ is taken 
from Ref.~\cite{ambainis2001}.}
\label{tab:right_all}
\end{table}

Next we express the right-half contribution to $\langle x^2 \rangle$ 
in terms of the $h_m$ defined in Section~IV (Eqs. (\ref{hm exact}),(\ref{apb}), 
and (\ref{amb})) and the Parseval identity (Eq. (\ref{Pq1})):
\be
&&\langle x^2 \rangle_{x\ge 0}=\frac{1}{2^n}\Big\{n^2+\frac{1}{2}
\sum^{[n/2]}_{k=1}(n-2 k)^2
\left[\left(K^{(n)}_{k,k-1}\right)^2+\left(K^{(n)}_{k,k}\right)^2\right]\Big\}
 \non \\
&=&\frac{n^2}{2^n}+ \sum^{[n/2]}_{k=1}\frac{(n-2 k)^2}{2^{n+1}}
\left\{\left[h_{k-1}(n-k+1,k)\binom{n-k+1}{k}\right]^2
+\left[h_{k}(n-k,k)\binom{n-k}{k}\right]^2
\right\}.\non \\\label{eq:var_qrw}
\en
In marked contrast, the RW 
gives 
\be
\langle x^2 \rangle_{\text{RW},x\ge 0} = \frac{1}{2^n}\sum_{k=0}^{[n/2]}
(n-2 k)^2\binom{n}{k}.\label{eq:var_rw}
\en
The two expressions differ structurally: the QRW variance involves 
squared amplitudes built from the $h_m$, whereas the RW 
variance is linear in the binomial coefficient. Nevertheless, as 
shown in Section~II, removing the alternating signs $(-1)^j$ from 
the definitions of $a$ and $b$ eliminates quantum interference and 
reduces the QRW distribution to the classical one.
Table~\ref{tab:var} shows, for $n=10$, the detailed breakdown of the 
right-half contributions to the variance. For each $k=1 \sim 5$, we 
list the quantities $h_{k-1}$ and $h_k$ that enter the two Krawtchouk 
matrix elements $K^{(10)}_{k,k-1}$ and $K^{(10)}_{k,k}$, together with 
the corresponding binomial coefficients and their normalized 
contributions to $\langle x^2 \rangle$. The last two columns compare 
the cumulative QRW and RW variances across the right half 
of the distribution. At $k=0$ ($x=10$), both the QRW and the 
RW contribute $100/2^{10} \approx 0.10$ to $\langle x^2 \rangle$, 
with no involvement of $h_m$.
\begin{table}[h]
\centering
\begin{tabular}{c|c|c|c|c|c|c|c|c|c}
\hline\hline
$k$ & $x$ &$h_{k-1}(11-k,k)$ &$\binom{11-k}{k}$ 
& $\left(x K^{(10)}_{k,k-1}\right)^2/2^{11}$ & $h_{k}(10-k,k)$ & $\binom{10-k}{k}$ 
& $\left(x K^{(10)}_{k,k}\right)^2/2^{11}$ & $\langle x^2 \rangle_{\text{QRW}}$ &  
$\langle x^2 \rangle_{\text{RW}}$
 \\\hline
1  & 8 & 1 & 10 & ~3.13  & 8/9  & 9& 2.00 & 5.13  & 0.63\\
2 & 6 &3/4& 36 &\!12.81 & 13/28 & 28 & 2.97 & 15.78 & 1.58\\ 
3 & 4 & 1/7 & 56 & 0.50&-8/35 & 35&0.50 &1.00 &1.88\\
4 & 2 &-2/5 &35&0.38& 2/15 &15 & 0.01 &0.39 &0.82\\
5& 0 & 2 & 6 &0&0 &1&0  & 0&0\\
\hline\hline
\end{tabular}
\caption{Right-half contributions to $\langle x^2 \rangle$ for the QRW with 
$n=10$. The quantities $h_m$ and the Krawtchouk matrix elements 
$K^{(10)}_{k,k-1}$, $K^{(10)}_{k,k}$ are evaluated at each position $k$. 
The last two columns compare the QRW and the RW contributions to the 
variance.}
\label{tab:var}
\end{table}
The numerical trends admit a straightforward interpretation in terms of the 
quantities $h_m$. For positions well inside the right half ($k =1,2$), the outer 
factor $(n-k)!/(n-k+m)!$ in Eq.~(\ref{hm exact}) decreases monotonically with 
$m$, providing an overall suppression of higher-order contributions. At the 
same time, the alternating sum in the numerator---inherited from the $(-1)^j$ 
factors---causes $h_m$ to oscillate as a function of $m$ and $k$. The 
combination of these two effects shifts the dominant contribution to 
$\langle x^2 \rangle$ away from the central region $x \approx 0$ 
and toward the ballistic front $|x| \approx n$, where $m$ is small 
and the outer factor is close to unity. Near the boundary $k \approx n/2$, the $h_m$ 
may deviate from this monotonic trend; however, this has a negligible 
effect on the variance because $x$ is small in that region. This mechanism underlies 
the $O(n^2)$ scaling of the variance. It is the discrete counterpart of the Hermite-
function oscillations that shape the QRW probability density in the continuous limit; 
the $O(n^2)$ scaling itself arises from the superposition of all 
excited states, not from any single Hermite function. The contrast is evident in the 
last two columns of Table~\ref{tab:var}: for $k=1,~2$, the QRW contributions already 
exceed their classical counterparts by a factor of $8~$--$~10$, while 
near the center the two become comparable, reflecting the 
suppression of quantum interference in that region. The ratio $\langle x^2 \rangle 
/ n^2$ provides a clean measure of this ballistic spreading. For 
$n=6$ it takes the value $0.313$, and by $n=10$ it has already 
dropped to $0.300$, approaching the asymptotic limit 
$1-1/\sqrt{2} \approx 0.293$ established in Ref.~\cite{ambainis2001}.

Finally we express the left-half contribution to $\langle x^2 \rangle$ 
in terms of the $h_m$:
\be
&&\langle x^2 \rangle_{x< 0}=\frac{1}{2^{n+1}}\sum^{n}_{k=[n/2]+1}(n-2 k)^2
\left[\left(K^{(n)}_{k,k-1}\right)^2+\left(K^{(n)}_{k,k}\right)^2\right] \non \\
&=& \sum^{n}_{k=[n/2]+1}\frac{(n-2 k)^2}{2^{n+1}}
\Big\{\left[\sum^{2k-n-1}_{l=0}(-1)^l\binom{2 k-n-1}{l}K^{(2 k-1-l)}_{k-l,k-1}\right]^2 \non\\
&&\qquad\qquad\qquad\qquad+\left[\sum^{2k-n}_{l=0}(-1)^l
\binom{2 k-n}{l}K^{(2 k-l)}_{k-l,k}\right]^2\Big\}\non \\
&=& \sum^{n}_{k=[n/2]+1}\frac{(n-2 k)^2}{2^{n+1}}
\Big\{\left[\sum^{2k-n-1}_{l=0}(-1)^l\binom{2 k-n-1}{l}h_{k-1}(k-l,k-l)\binom{k-l}{k-l}\right]^2 \non \\
&&\quad\quad\quad\quad\quad\quad\quad\quad+\left[\sum^{2k-n}_{l=0}(-1)^l
\binom{2 k-n}{l}h_{k}(k-l,k-l)\binom{k-l}{k-l}\right]^2\Big\}. \label{eq:var_qrws0}
\en
Once again the inverse Pascal triangle structure appears, but this 
time the Krawtchouk matrix elements involved range from $K^{(n+1)}$ up to 
$K^{(2n)}$, depending on the number of terms in each alternating sum. 
Table \ref{tab:var2} shows the left-half contributions to $\langle x^2 \rangle$ 
for the QRW with $n=10$, we list the squared 
values of $K^{(n)}_{k,k-1}$ and $K^{(n)}_{k,k}$ for $k = 6$ to $10$.
\begin{table}[h]
\centering
\begin{tabular}{c|c|c|c|c|c|c|c|c}
\hline\hline
$k$ & $x$ &$x^2$ &$\left(K^{(10)}_{k,k-1}\right)^2$ 
& $\left(x K^{(10)}_{k,k-1}\right)^2/2^{11}$ & $\left(K^{(10)}_{k,k}\right)^2$ 
& $\left(x K^{(10)}_{k,k}\right)^2/2^{11}$ & $\langle x^2 \rangle_{\text{QRW}}$ &  
$\langle x^2 \rangle_{\text{RW}}$
 \\\hline
6  & -2 &4 & 100 & 0.20  & 4  & 0.01 & 0.20  & 0.82\\
7 & -4 &16& 64 & 0.50 & 64  & 0.50 & 1.00 & 1.88\\ 
8 & -6 & 36 & 9 & 0.16&169 &2.97 &3.13&1.58\\
9 & -8 &64 &36&1.13& 64& 2.00 &3.13 &0.63\\
10&-10 &100& 1 &0.05&1 &0.05  & 0.10&0.10\\
\hline\hline
\end{tabular}
\caption{Left-half contributions to $\langle x^2 \rangle$ for the QRW with $n=10$. 
The Krawtchouk matrix elements $K^{(10)}_{k,k-1}$, $K^{(10)}_{k,k}$ 
are evaluated at each position $k$. The last two columns compare the QRW and the RW 
contributions to the variance.}
\label{tab:var2}
\end{table}
As discussed above, the combination of several $h_m$ in the form of 
a difference structure makes the probability distribution for 
$x < 0$ less sharp. This also makes the contributions to $\langle x^2 \rangle$ in the 
$x < 0$ region smaller and more evenly distributed; nevertheless, 
at the relatively large $k = 8, 9$ positions, their magnitudes 
remain $2$--$5$ times larger than those of the RW 
distribution. Hence the ballistic spreading of the QRW persists 
across the entire distribution, with the left half contributing 
a smaller but still significant share of the variance. In summary, the overall ballistic 
spreading of the QRW does not arise from a single discrete Hermite function, 
but from a position-dependent superposition of discrete Hermite 
functions, with different numbers of them contributing at 
different locations---either individually (for $x \ge 0$) or in 
inverse-Pascal combinations (for $x < 0$).
\section{Conclusions}
This work introduced a difference operator $D_k$ to study the one-dimension QRW. 
We first derived the probability amplitudes $a(n,k)$ and $b(n,k)$ for the 
Hadamard walk with the initial state $|R\ra\otimes|0\ra$. The inverse Pascal triangle 
structure of $a$ and $b$ reveals not only the convolution of phase factor and binomial
coefficients, but also the path by which the RW was recovered, we show
that, when the phase factors responsible for interference and the coin states encoding
quantum memory are discarded, the classical walk emerges.

Next, we connected the amplitudes $a(n,k)$ and $b(n,k)$ to the Krawtchouk matrix via 
generating functions. Using the cross identities among the Krawtchouk matrix elements,
we proved by induction that these probability amplitudes satisfy a Parseval identity, i.e., 
probability is conserved. This result shows that the unitary character of  the QRW survives
in the Krawtchouk formulation.

The inverse Pascal triangle structure of $a$ and $b$ also motivated the introduction of
a discrete difference operator $D_k$, which ties together the probability amplitudes, 
binomial coefficients, and the Krawtchouk matrix together. In analogy with 
the Rodrigues formula for Hermite polynomials, $D_k$ acts as a generator for a discrete 
counterpart  $h_m(n,k)$ of the Hermite polynomial $H_m(\xi)$. The convergence of $h_m$ 
and $H_m$ in the continuous limit is consistent with the Askey limit of the Krawtchouk
polynomicals.

In the last part of this study, we used the $n=6$ example to illustrate 
the relationships among the amplitudes $a$ and $b$, the Krawtchouk 
matrix elements, the classical binomial distribution, the difference 
operator $D_k$, and the quantities $h_m$. In the right half ($x \ge 0$), 
the symmetric and antisymmetric combinations $a+b$ and $a-b$ 
correspond respectively to the diagonal and left-adjacent sub-diagonal 
entries of the Krawtchouk matrix. Each of these arises from repeated 
applications of $D_k$---acting as a raising operator---to a single 
classical binomial distribution at a smaller value of $n$, and is governed 
by a single $h_m$. The resulting $h_m$ quantify the degree of quantum 
interference produced by $D_k$, which thereby acts as a coherence 
generator; different positions in the QRW experience different degrees 
of quantum interference. In the left half ($x < 0$), the Krawtchouk matrix 
elements are instead coherent combinations of several $h_m$, weighted 
by the inverse Pascal triangle. The alternating signs inherent in this 
difference structure lead to partial cancellations that smooth out 
the distribution, making the left peak broader and lower than its 
right counterpart. 

The ballistic spreading persists across both halves: as shown in 
Table~\ref{tab:var} and \ref{tab:var2}, the QRW contributions to 
$\langle x^2 \rangle$ in the right and left half remain $8$--$10$ 
and $2$--$5$ times larger than those of the RW, respectively. 
It is precisely this position-dependent interference strength---
governed by $h_m$ and by the inverse Pascal structure---that 
accounts for the ballistic spreading. Hence, the overall behavior 
does not arise from a single discrete Hermite function, but from 
a position-dependent superposition of them, with different numbers 
contributing at different locations.

Several directions remain open for future investigation. First, the 
difference operator $D_k$ introduced here is specifically tailored to 
the Hadamard walk with the initial state $|R\rangle \otimes |0\rangle$. 
Extending this framework to other coin operators and to symmetric 
initial states would clarify the full scope of the combinatorial 
approach. Second, the discrete continuity equation hinted at by the Parseval 
identity deserves a systematic derivation, which would clarify the 
local mechanism of probability transport in the QRW.
Finally, the generalization to higher-dimensional QRWs via 
multivariate Krawtchouk products offers a natural extension, where 
the difference operator would split into components associated with 
each spatial direction. 

{\bf Acknowledgements}\\
I would like to thank Dr. Kao Chan-Peng for useful discussions.

\appendix
\section{Derivation of $a-b = K_{k,k}^{(n)}$ via Generating Functions}
\label{app:genfun}
From the binomial theorem, the two polynomials can be expanding as
 \be
(1- z)^{k-1} &=& \sum_{l=0}^{k-1} (-1)^l \binom{k-1}{l} z^l,\\
(1+z)^{n-k} &=& \sum_{m=0}^{n-k} \binom{n-k}{m} z^m.
 \en
Now we consider the product of these two polynomials:
 \be
F_{n,k}(z) = (1-z)^{k-1} (1+z)^{n-k}=\left( \sum_{j=0}^{k-1} (-1)^j \binom{k-1}{j} 
z^j \right) \cdot \left( \sum_{m=0}^{n-k} \binom{n-k}{m} z^m \right).
 \label{A2}
 \en
The coefficient of $z^k$ in the product comes from all terms that satisfy 
$l + m = k$, i.e., $m = k - l$:
 \be
 [z^k] F_{n,k}(z) = \sum_{l=0}^{k-1} (-1)^l \binom{k-1}{l} \binom{n-k}{k-l},
 \label{A3}
 \en
and equals to $a(n,k)$ exactly. Thus, 
\be
a(n,k)=[z^k](1-z)^{k-1}(1+z)^{n-k}.
\en
Likewise, the coefficient of $z^{k-1}$ in the product equals to $b(n,k)$
\be
b(n,k)=[z^{k-1}](1-z)^{k-1}(1+z)^{n-k}.
\en
From the relation: $[z^{k-1}]P(z)=[z^{k}](z \cdot P(z))$, we obtain
\be
b(n,k)=[z^{k}]\left(z\cdot (1-z)^{k-1}(1+z)^{n-k}\right).
\en
Therefore, the generating function of $a(n,k)-b(n,k)$ is
\be
a(n,k)-b(n,k)&=&[z^{k}]\left((1-z)^{k-1}(1+z)^{n-k}-z(1-z)^{k-1}(1+z)^{n-k}\right)\non\\
&=&[z^{k}](1-z)^{k-1}(1+z)^{n-k}(1-z)\non\\
&=&[z^{k}](1-z)^{k}(1+z)^{n-k}. \label{A8}
\en
Comparing Eq. (\ref{A8}) with Eq.(\ref{Kdef}), we finally obtain
\be
a(n,k)-b(n,k)=K^{n}_{k.k}.
\en


\begin{thebibliography}{99}

\bi{childs2009}
A. M.~Childs, \prl, \textbf{102},  180501 (2009).

\bi{lovett2010}
N. B.~Lovett, S.~Cooper, M.~Everitt, M.~Trevers, and V.~Kendon, 
Phys.\ Rev.\ A  \textbf{81}, 042330 (2010).

\bibitem{aharonov1993}
Y.~Aharonov, L.~Davidovich, and N.~Zagury,
Phys.\ Rev.\ A \textbf{48}, 1687 (1993).

\bibitem{ambainis2001}
A.~Ambainis, E.~Bach, A.~Nayak, A.~Vishwanath, and J.~Watrous,
Proceedings of the 33rd Annual ACM Symposium on Theory of Computing, 37--49 
(2001).

\bibitem{kempe2003}
J.~Kempe,
Contemporary Physics \textbf{44}, 307--327 (2003).

\bi{DTQW}
D.~Aharonov, A.~Ambainis, J.~Kempe, and U.~Vazirani, Proceedings of the 33rd 
Annual ACM Symposium on Theory of Computation, 50--59 (2001).

\bi{CTQW}
E.~Farhi  S.~Gutmann, Phys.\ Rev.\ A \textbf{58}, 915 (1998).

\bi{faster}
A.~Ambainis, J.~Kempe, and A.~Rivosh,
Proceedings of the 16th ACM-SIAM SODA, 1099 (2005).

\bi{meyer1996}
D. A.~Meyer, 
J. Stat. Phys. \textbf{85}, 551-574 (1996).

\bi{grimmett2004}
G.~Grimmett, S.~Janson, and P. F.~Scudo,
Phys.\ Rev.\ E \textbf{69}, 026119 (2004).

\bi{konno2002}
N.~Konno,
Quantum Information Processing, \textbf{1}, 345 (2002).

\bibitem{feinsilver2004}
P.~Feinsilver and J.~Kocik,
in {\em Recent Advances in Applied Probability}, Springer, 115--141 (2004).

\bibitem{feinsilver2007}
P.~Feinsilver and J.~Kocik,
Contemporary Mathematics \textbf{287}, 83-96 (2001).


\bi{askey1985}
R.~Askey and J.~Wilson,  Memoirs of AMS,  \textbf{54}, 319 (1985).

\bi{koekoek2010}
R.~Koekoek, P. A.~Lesky, and R. F.~Swarttouw, Hypergeometric orthogonal 
polynomials and their q-analogues. Springer (2010).



\end{thebibliography}
\end{document}